\def\bar{\overline}

\def\a{\alpha}
\def\b{\beta}

\def\ee{e}

\def\th{\theta}

\def\bar{\overline}

\def\eV{{\rm eV}}
\documentstyle [12pt,epsf]{article}
\begin{document}
\baselineskip=18  pt
\setcounter{page}{1}
\thispagestyle{empty}
\topskip 0.5  cm
\vspace{1 cm}
\centerline{\LARGE \bf Impact of $U_{\ee3}$ on Neutrino Models 
\footnote{Invited talk presented at NOON2003 (February 2003)
 at Kanazawa, Japan.}
}
\vskip 0.5 cm
\centerline{{\large \bf Morimitsu Tanimoto}
\renewcommand{\thefootnote}{\fnsymbol{footnote}}
\footnote[1]{E-mail address: tanimoto@muse.sc.niigata-u.ac.jp}
 }
\vskip 0.5 cm
 \centerline{ \it{Department of Physics, Niigata University, 
 Ikarashi 2-8050, 950-2181 Niigata, JAPAN}}
\vskip 1 cm
\centerline{\bf ABSTRACT}\par
\vskip 0.3 cm
We have discussed the impact of  $U_{e3}$ on the 
model of the neutrino mass matrix. 
In order to get the small  $U_{e3}$,  some flavor symmetry
is required. Typical two models  are investigated. The first one is the
model  in which the bi-maximal mixing is realized at  the symmetric limit.
The second one is the texture zeros of the neutrino mass matrix.

\section{Introduction}
 In these years empirical understanding of the mass and mixing of 
 neutrinos  have been  advanced \cite{SKam,SKamsolar,SNO}.
The KamLAND experiment selected the neutrino
mixing solution that is responsible for the solar neutrino problem
nearly uniquely \cite{KamLAND},   only large mixing angle solution.
We have now good understanding concerning the neutrino
mass difference squared  and 
neutrino flavor mixings \cite{Lisi}. A  constraint has also
been placed on the mixing from the reactor experiment of CHOOZ \cite{CHOOZ}.

 These results indicate two large flavor mixings and  one small
flavor mixing. 
It is therefore important to investigate how the textures of lepton mass 
matrices can link up with the observables of the flavor mixing.
There are some ideas to explain the large mixing angles.
The mass matrices, which lead to the large mixing angle,
are ``lopsided mass matrix''\cite{SY}, ``democratic mass matrix''
\cite{FTY}
and ``Zee  mass matrix''\cite{Zee}. 
These textures are reconciled with some flavor symmetry.

We have another problem. Is the small $U_{e3}$ always guaranteed
 in the model with two large mixing angles?
The answer is ``No''. There are some models to give a large $U_{e3}$.
The typical one is ``Anarchy'' mass matrix \cite{AN}, which gives a rather
large  $U_{e3}$.
Another example is the model, in which
the large  solar neutrino mixing comes from the charged lepton sector
while  the large atmospheric neutrino mixing comes from the neutrino
sector. In this model   $U_{e3}=1/2$ is predicted.

In order to get the small  $U_{e3}$,  some flavor symmetry is required.
Typical two models  are investigated in this talk. The first one is the
model  in which the bi-maximal mixing is realized at  the symmetric limit.
The second one is the texture zeros of the neutrino mass matrix.

\section{Deviation from the Bi-Maximal Mixings}

We consider the  symmetric limit with the bi-maximal flavor mixing 
at which $U_{e3}=0$ as follows \cite{GT}:
\begin{eqnarray}
   \nu_\a  =  U^{(0)}_{\a i} \nu_i \ ,
\end{eqnarray}
where
\begin{eqnarray}
U^{(0)}= \left ( \matrix{\frac{1}{\sqrt{2}}& \frac{1}{\sqrt{2}} & 0\cr
                -\frac{1}{2} &  \frac{1}{2}& \frac{1}{\sqrt{2}} \cr
 \frac{1}{2} & -\frac{1}{2} &  \frac{1}{\sqrt{2}} \cr } \right ) \ .
\end{eqnarray}
\noindent One  can parametrize the deviation  $U^{(1)}$ in
$\nu_\a =  [ U^{(1)^\dagger} U^{(0)}]_{\a i} \nu_i$  as
follows:
\begin{equation}
U^{(1)} =\left (\matrix{c^1_{13}c^1_{12} & c^1_{13} s^1_{12} & s^1_{13}e^{-i
\phi}\cr
  -c^1_{23}s^1_{12}-s^1_{23}s^1_{13}c^1_{12}e^{i \phi} &
c^1_{23}c^1_{12}-s^1_{23}s^1_{13}s^1_{12}e^{i \phi} &
                       s^1_{23}c^1_{13} \cr
  s^1_{23}s^1_{12}-c^1_{23}s^1_{13}c^1_{12}e^{i \phi} &
-s^1_{23}c^1_{12}-c^1_{23}s^1_{13}s^1_{12}e^{i \phi} &
                       c^1_{23}c^1_{13} \cr} \right )
\label{Mix}
\end{equation}
where  $s^1_{ij}\equiv \sin{\theta^1_{ij}}$ and $c^1_{ij}\equiv
\cos{\theta^1_{ij}}$ denote the mixing angles  in the bi-maximal basis and
$\phi$ is the CP violating Dirac phase.
The mixings $s^1_{ij}$ are expected to be small since these are deviations
from the bi-maximal mixing.
Here, the Majorana phases are absorbed in the neutrino mass eigenvalues.

Let us assume the mixings  $s^1_{ij}$ to be  hierarchical like the ones
 in the quark sector, $s^1_{12}\gg s^1_{23}\gg s^1_{13}$.
Then, taking the leading contribution due to $s^1_{12}$, we have
\begin{eqnarray}
|U_{e1}|\simeq \frac{1}{\sqrt{2}} \left (
c^1_{12}+\frac{1}{\sqrt{2}}s^1_{12} \right ) \ , \quad
|U_{e2}|\simeq \frac{1}{\sqrt{2}} \left (
c^1_{12}-\frac{1}{\sqrt{2}}s^1_{12} \right ) \ , 
\end{eqnarray}
which lead to
\begin{equation}
\tan^2 \theta_{\rm sol}
=
\left( \frac{ c^1_{12} - \frac{1}{\sqrt{2}}s^1_{12} }{ c^1_{12} +
\frac{1}{\sqrt{2}}s^1_{12} } \right)^2
=
1 - 2 \sqrt{2} s^1_{12}
+ \mathrm{O}({(s^1_{12})}^2)
\,.
\end{equation}
Thus, the solar neutrino mixing is somewhat reduced due to $s^1_{12}$.
By using the data of the solar neutrino mixing, 
we predict the small $U_{e3}$ such as 
\begin{equation}
|U_{e3}|\simeq  \frac{1}{\sqrt{2}} s^1_{12} = 0.03 \sim 0.2
\ ,
\end{equation}
\noindent which is testable in the future experiments.
In the next section, we present another approach, texture zeros.

\section{Texture Zeros of Neutrino Mass Matrix}
The texture zeros of the neutrino mass matrix have been discussed
 to explain these  neutrino masses and mixings  \cite{Fu,AK,Kang}.
  Recently, Frampton, Glashow and Marfatia \cite{Fram} found 
acceptable textures of the  neutrino mass matrix with two independent
vanishing entries in the basis of the diagonal charged lepton mass matrix.
 The  KamLAND result has stimulated the phenomenological analyses
of the texture zeros
\cite{Xing1,Xing2,Barbieri,Obara}.
These results favour texture zeros for the neutrino mass matrix
phenomenologically.

 There are  15 textures with two zeros  for the effective neutrino mass matrix $M_\nu$, which have five independent parameters. 
The two zero conditions give
\begin{equation}
 (M_\nu)_{ab}=\sum^3_{i=1} U_{ai} U_{bi}\lambda_i = 0 \ , \qquad\qquad
 (M_\nu)_{\a\b}=\sum^3_{i=1} U_{\alpha i} U_{\beta i}\lambda_i = 0 \ ,
\label{cons1} 
\end{equation}
\noindent  where $\lambda_i$ is the i-th eigenvalue
including the  Majorana phase, and indices $(a b)$ and $ (\a  \b)$ denote 
 the  flavor components, respectively.

Solving these equations, the ratios of neutrino masses $m_1$, $m_2$, $m_3$,
 which are absolute values of $\lambda_i$'s, are given
in terms of the neutrino mixing matrix  $U$ \cite{MNS} as follows:
\begin{eqnarray} 
\frac{m_1}{m_3} & = & \left |
\frac{U_{a3} U_{b3} U_{\alpha 2} U_{\beta 2} - U_{a2} U_{b2} U_{\alpha 3}
U_{\beta 3}}{U_{a2} U_{b2} U_{\alpha 1} U_{\beta 1} - U_{a1} U_{b1}
U_{\alpha 2} U_{\beta 2}} \right | \ ,
\nonumber \\ 
\frac{m_2}{m_3} & = & \left |
\frac{U_{a1} U_{b1} U_{\alpha 3} U_{\beta 3} - U_{a3} U_{b3} U_{\alpha 1}
U_{\beta 1}}{U_{a2} U_{b2} U_{\alpha 1} U_{\beta 1} - U_{a1} U_{b1}
U_{\alpha 2} U_{\beta 2}} \right |  \ .
\end{eqnarray}
\noindent Then, one can test textures in the ratio $R_\nu$,
\begin{eqnarray}
R_\nu \equiv \left | \frac{m^2_2 - m^2_1}
{m^2_3 - m^2_2} \right | \approx \frac{\Delta m^2_{\rm sun}}
{\Delta m^2_{\rm atm}} \  ,
\label{Rnu}
\end{eqnarray}
\noindent which has been  given by the experimental data.
 The ratio $R_\nu$ is given only in terms of four parameters 
 (three mixing angles and one phase) in 
\begin{eqnarray}
 U = \left (\matrix{ c_{13} c_{12} & c_{13} s_{12} &  s_{13} e^{-i \delta}\cr 
  -c_{23}s_{12}-s_{23}s_{13}c_{12}e^{i \delta}
 & c_{23}c_{12}-s_{23}s_{13}s_{12}e^{i \delta} &   s_{23}c_{13} \cr
  s_{23}s_{12}-c_{23}s_{13}c_{12}e^{i \delta} 
& -s_{23}c_{12}-c_{23}s_{13}s_{12}e^{i \delta} & c_{23}c_{13} \cr}\right )
\end{eqnarray}
\noindent where $c_{ij}$ and  $s_{ij}$ denote
 $\cos \theta_{ij}$ and $\sin \theta_{ij}$, respectively.

 Seven acceptable textures with two independent 
zeros were found for the  neutrino mass matrix \cite{Fram}, 
and they have been studied in detail
\cite{Xing2,Barbieri}.
Among them, the textures $\rm A_1$ and $\rm A_2$ \cite{Fram}, which  
correspond to the hierarchical neutrino mass spectrum, are strongly favoured 
 by the recent phenomenological analyses \cite{Xing1,Xing2,Barbieri}.
  Therefore,  we study these two  textures in this paper.

In the texture $A_1$,
 which has two zeros as  $(M_\nu)_{ee}=0$ and  $(M_\nu)_{e\mu}=0$, 
the mass ratios are given as
\begin{eqnarray}
\frac{m_1}{m_3} & = &
 \left | \frac{s_{13}}{c^2_{13}} \left (\frac{s_{12} s_{23}}{c_{12} c_{23}}-
 s_{13}  e^{-i\delta}\right ) \right |\ ,
\nonumber \\
\frac{m_2}{m_3} & = &
 \left |\frac{s_{13}}{c^2_{13}} \left ( \frac{c_{12} s_{23}}{s_{12} c_{23}} +
 s_{13}  e^{-i\delta}\right )\right | \ .
\end{eqnarray}
\noindent In  the texture $A_2$,
 which has two zeros as  $(M_\nu)_{ee}=0$ and  $(M_\nu)_{e\tau}=0$, 
the mass ratios are given as
\begin{eqnarray}
\frac{m_1}{m_3} & = &
 \left | \frac{s_{13}}{c^2_{13}} \left (\frac{s_{12} c_{23}}{c_{12} s_{23}}+
 s_{13}  e^{-i\delta}\right ) \right | \ ,
\nonumber \\
\frac{m_2}{m_3} & = &
 \left |\frac{s_{13}}{c^2_{13}} \left ( \frac{c_{12} c_{23}}{s_{12} s_{23}} -
 s_{13}  e^{-i\delta}\right )\right | \ .
\end{eqnarray}
\noindent
 If $\theta_{12}$, $\theta_{23}$,  $\theta_{13}$ and $\delta$ are fixed,
 we can predict $R_\nu$ in eq.(\ref{Rnu}), which  can be  compared with the experimental
value ${\Delta m^2_{\rm sun}}/{\Delta m^2_{\rm atm}}$.

Taking account of the following data with $90\%$ C.L. \cite{Lisi},
\begin{eqnarray}
 \sin^2 2\th_{\rm atm} \geq 0.92\ , \qquad
 &&\Delta m^2_{\rm atm}=  (1.5\sim 3.9)\times  10^{-3} \eV^2\ , \nonumber \\
 \tan^2 \th_{\rm sun}=0.33\sim 0.67\ ,  \qquad
 &&\Delta m_{\rm sun}^2= (6\sim 8.5)\times 10^{-5}\eV^2\ ,
\label{Data}
\end{eqnarray}
\noindent with  $\sin \th_{\scriptscriptstyle \rm CHOOZ} \leq 0.2$, 
we predict $R_\nu$.
\begin{figure}
\epsfxsize=7.0 cm
\centerline{\epsfbox{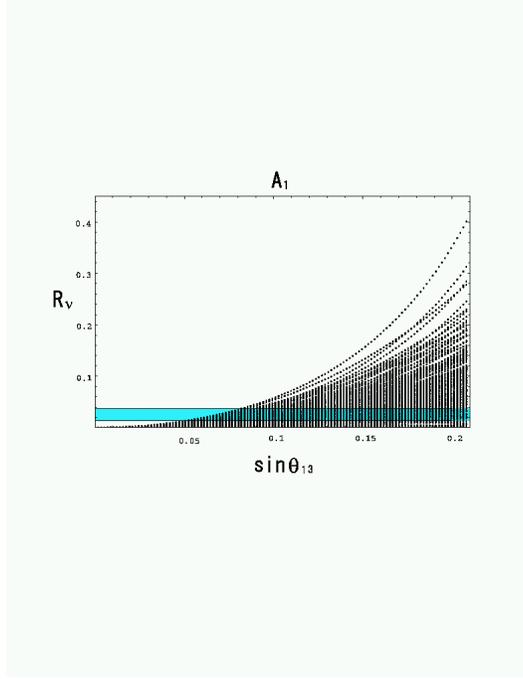}}
\caption{ Scatter plot of $R_\nu$  versus $\sin\theta_{13}$ for the texture
$A_1$. The unknown phase $\delta$ is taken in the whole region $-\pi\sim \pi$.
 The gray horizontal band is the experimental allowed region.}
\end{figure}
In Fig.1,  we present the scatter plot of  the predicted $R_\nu$ versus 
$\sin\theta_{13}$,
in which $\delta$ is taken in the whole range $-\pi \sim \pi$ 
for the texture $A_1$.
 The parameters are taken in the following ranges in 
 $\theta_{12}=30^\circ\sim 39^\circ$, $\theta_{23}=37^\circ\sim 53^\circ$,
 $\theta_{13}=1^\circ\sim 12^\circ$ and $\delta=-\pi\sim\pi$ 
with constant distributions those are flat on a linear scale.
It is found that 
many predicted values of $R_\nu$ lie outside the experimental allowed region. 
This result means  that  some tunings among four  parameters are demanded
 to be consistent with the experimental data.
We get $\sin\theta_{13}\geq 0.05$ from
the experimental value  of $R_\nu$ as seen in Fig.1. 

In order to present the allowed  region of $\sin\theta_{13}$,
 we show the scatter plot  of $\sin\theta_{13}$ versus
$\tan^2\theta_{12}$ and $\tan^2\theta_{23}$ in Fig.2 and Fig.3, respectively,
for the texture $A_1$. 
 For the texture  $A_2$, the numerical results is similar with the one 
in the texture  $A_1$ because those are obtained only by replacing 
$\tan\theta_{23}$ in $A_1$ with  $-\cot\theta_{23}$.
 
The allowed regions in Fig.2 and Fig.3 are  quantitatively  understandable
in the following approximate relations:
\begin{eqnarray}
 |U_{e3}|\equiv\sin\theta_{13}\simeq
  \frac{1}{2}\tan 2\theta_{12} \ \cot \theta_{23}
\sqrt{R_\nu \cos 2\theta_{12}} \ ,
\end{eqnarray}
\noindent for the texture $A_1$, and 
\begin{eqnarray}
 |U_{e3}|\equiv\sin\theta_{13}\simeq \frac{1}{2}\tan 2\theta_{12} \ \tan \theta_{23}
\sqrt{R_\nu \cos 2\theta_{12}} \ ,
\end{eqnarray}
\noindent for the texture $A_2$, respectively,
where the phase $\delta$  is neglected because it is a next leading term.
 As $\tan\theta_{12}$ increases, the lower bound of $|U_{e3}|$ increases,
and as  $\tan\theta_{23}$ decreases, it  increases.
It is found in Fig.2 that the lower bound  $|U_{e3}|=0.05$ is given
 in the case of   the smallest $\tan^2\theta_{12}$, while
 $|U_{e3}|=0.08$  is given in the largest  $\tan^2\theta_{12}$.
On the other hand, as seen in Fig.3, the lower bound $|U_{e3}|=0.05$ is given 
in the largest $\tan^2\theta_{23}$, while $|U_{e3}|=0.08$ is given
 in  the smallest  $\tan^2\theta_{23}$.
In the future,  error bars of experimental data in eq.(\ref{Data})
 will be reduced.  Especially, KamLAND is expected to determine
 $\Delta m^2_{12}$ precisely. Therefore, the predicted region of  $|U_{e3}|$ 
 will be reduced significantly in the near future.
\begin{figure}
\epsfxsize=7.5 cm
\centerline{\epsfbox{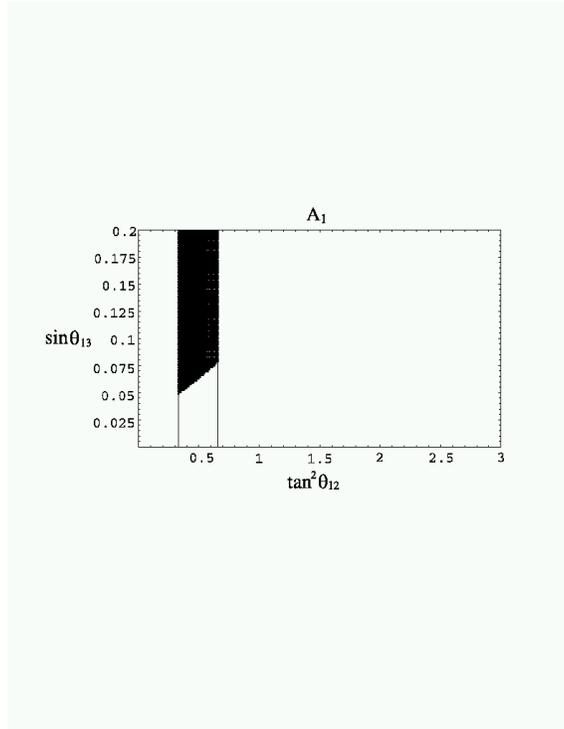}}
\caption{ Scatter plot of $\sin\theta_{13}$ versus $\tan^2\theta_{12}$
for the texture $A_1$. }
\end{figure}
\begin{figure}
\epsfxsize=7.5 cm
\centerline{\epsfbox{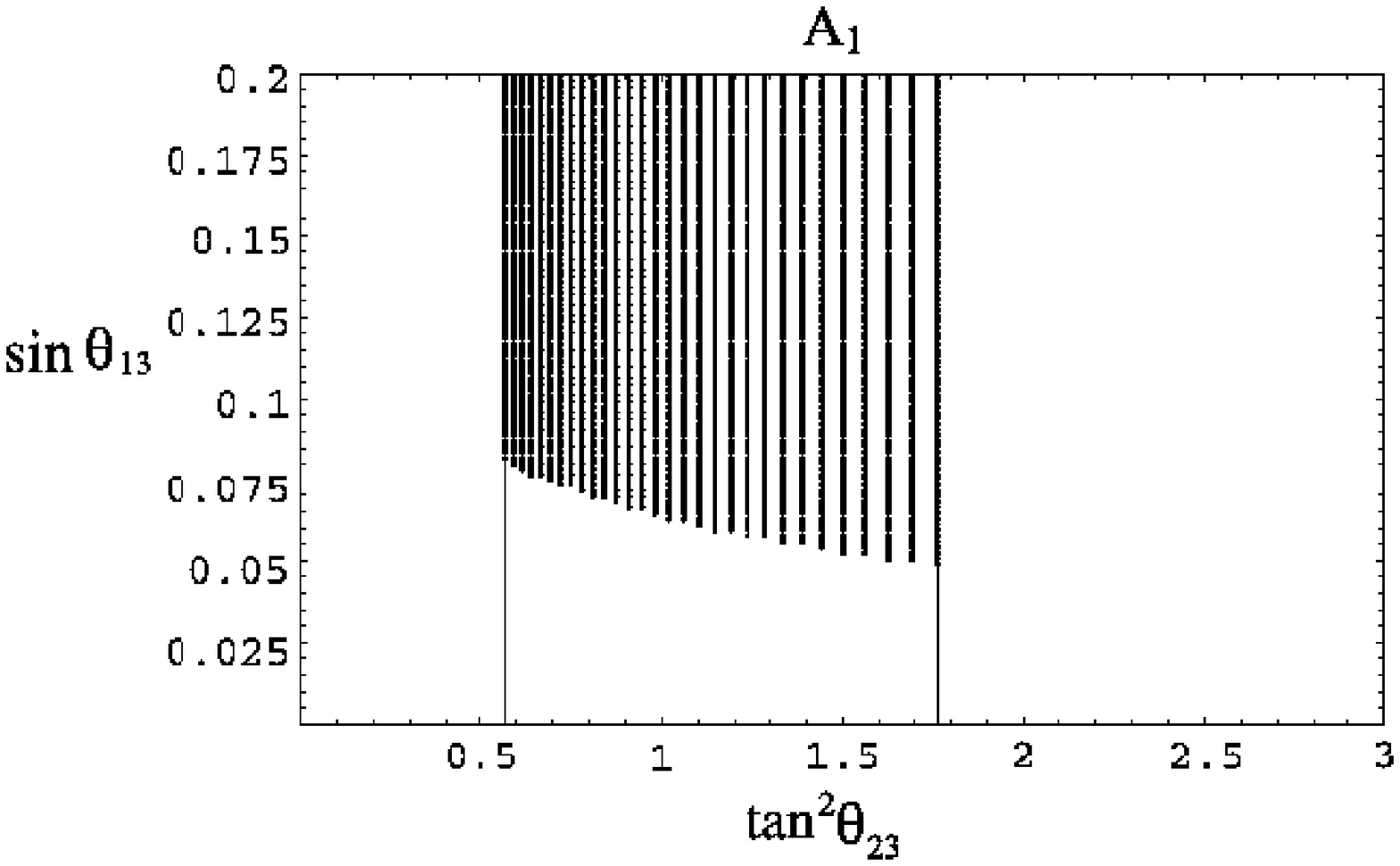}}
\caption{ Scatter plot of $\sin\theta_{13}$ versus $\tan^2\theta_{23}$
for the texture $A_1$.   }
\end{figure}

 Above  predictions are important ones in  the texture zeros.
The relative magnitude of each entry of the neutrino mass matrix is roughly
given as follows: 
\begin{eqnarray}
M_\nu \sim \left ( \matrix{
{ 0} & {0} & \lambda \cr {0} & 1 & 1 \cr \lambda  & 1 & 1 \cr} \right )
\  \ \ {\rm for\  A_1} \ , \qquad
\qquad 
\left (\matrix{ 0 & \lambda & 0\cr\lambda  & 1 & 1 \cr 0 & 1 & 1 \cr} \right )
\  \ \ {\rm for \ A_2} \ ,
\label{A12}
\end{eqnarray}
\noindent
where  $\lambda\simeq 0.2$.   
However,  these  texture zeros  are not  preserved to all orders. 
 Even if zero-entries of the mass matrix are given at the high energy scale,
 non-zero components may evolve instead of zeros at the low energy scale
  due to  radiative corrections.
Those magnitudes depend on unspecified interactions from which lepton masses
are generated. 
Moreover,  zeros  of the neutrino mass matrix are
given  while the charged lepton mass matrix has off-diagonal components
in the model with  some flavor symmetry. 
 Then, zeros are not  realized in the diagonal basis of the 
charged lepton mass matrix.
In other words, zeros of the neutrino mass matrix is  polluted by 
the small off-diagonal elements of the charged lepton mass matrix.

Therefore, one need the careful study of stability of the prediction for 
 $U_{e3}$ because this is a  small quantity.
In order to see the effect of the small non-zero components,
the conditions of zeros in  eq.(\ref{cons1}) are changed.
The two  conditions   turn to
\begin{equation}
 (M_\nu)_{ab}=\sum^3_{i=1} U_{ai} U_{bi}\lambda_i = \epsilon \ , \qquad\qquad
 (M_\nu)_{\a\b}=\sum^3_{i=1} U_{\alpha i} U_{\beta i}\lambda_i = \omega \ ,
\label{cons2} 
\end{equation}
\noindent
where $\epsilon$ and $\omega$ are arbitrary parameters with the mass unit,
 which are
much smaller than other non-zero components of the mass matrix.  
These parameters are supposed to be real for simplicity.
For the texture $A_1$,  we get
\begin{eqnarray} 
\frac{m_1}{m_3} & = & \left |
\frac{U_{13} U_{13} U_{12} U_{22} - U_{12} U_{12} U_{13}U_{23}
-U_{12} U_{22}\bar\epsilon + U_{12} U_{12}\bar\omega}
 {U_{12} U_{12} U_{11} U_{21} - U_{11} U_{11} U_{12} U_{22}} \right | \ ,
\nonumber \\ 
\frac{m_2}{m_3} & = & \left |
\frac{U_{11} U_{11} U_{13} U_{23} - U_{13} U_{13} U_{11}U_{21}
-U_{11} U_{21}\bar\epsilon + U_{11} U_{11}\bar\omega}
{U_{12} U_{12} U_{11} U_{21} - U_{11} U_{11}U_{12} U_{22}} \right |  \ ,
\label{ratio}
\end{eqnarray}
\noindent
where $\bar\epsilon$ and $\bar\omega$ are normalized ones as
$\bar\epsilon=\epsilon/\lambda_3$ and $\bar\omega=\omega/\lambda_3$, 
respectively.   We obtain approximately   
\begin{eqnarray}
\frac{m_1}{m_3} & \simeq &  s_{13} t_{12} t_{23}- \frac{t_{12}}{c_{23}}\bar\omega
+ \bar\epsilon \ ,
\nonumber \\
\frac{m_2}{m_3} & \simeq &  - s_{13}\frac{1}{t_{12}} t_{23}- \frac{1}{t_{12}c_{23}}\bar\omega - \bar\epsilon\ ,
 \end{eqnarray}
\noindent where $t_{ij}=\tan\theta_{ij}$.
The $|U_{e3}|=\sin\theta_{13}$ is given as
\begin{eqnarray}
 \sin\theta_{13}&\simeq&\frac{1}{2}\tan 2\theta_{12}\ \cot\theta_{23}
\sqrt{R_\nu \cos 2\theta_{12}} - \frac{\bar\omega}{\sin\theta_{23}}
\  \frac{1+\tan^4\theta_{12}}{1-\tan^4\theta_{12}} 
\nonumber\\
 && - \frac{t_{12}}{t_{23}}\frac{\bar\epsilon}{1+\tan^2\theta_{12}}\ .
\label{Ue3corr}
\end{eqnarray}
\noindent
It is remarked that the second and third terms in the right hand side could 
be comparable with the first one.
 \begin{figure}
\epsfxsize=7 cm
\centerline{\epsfbox{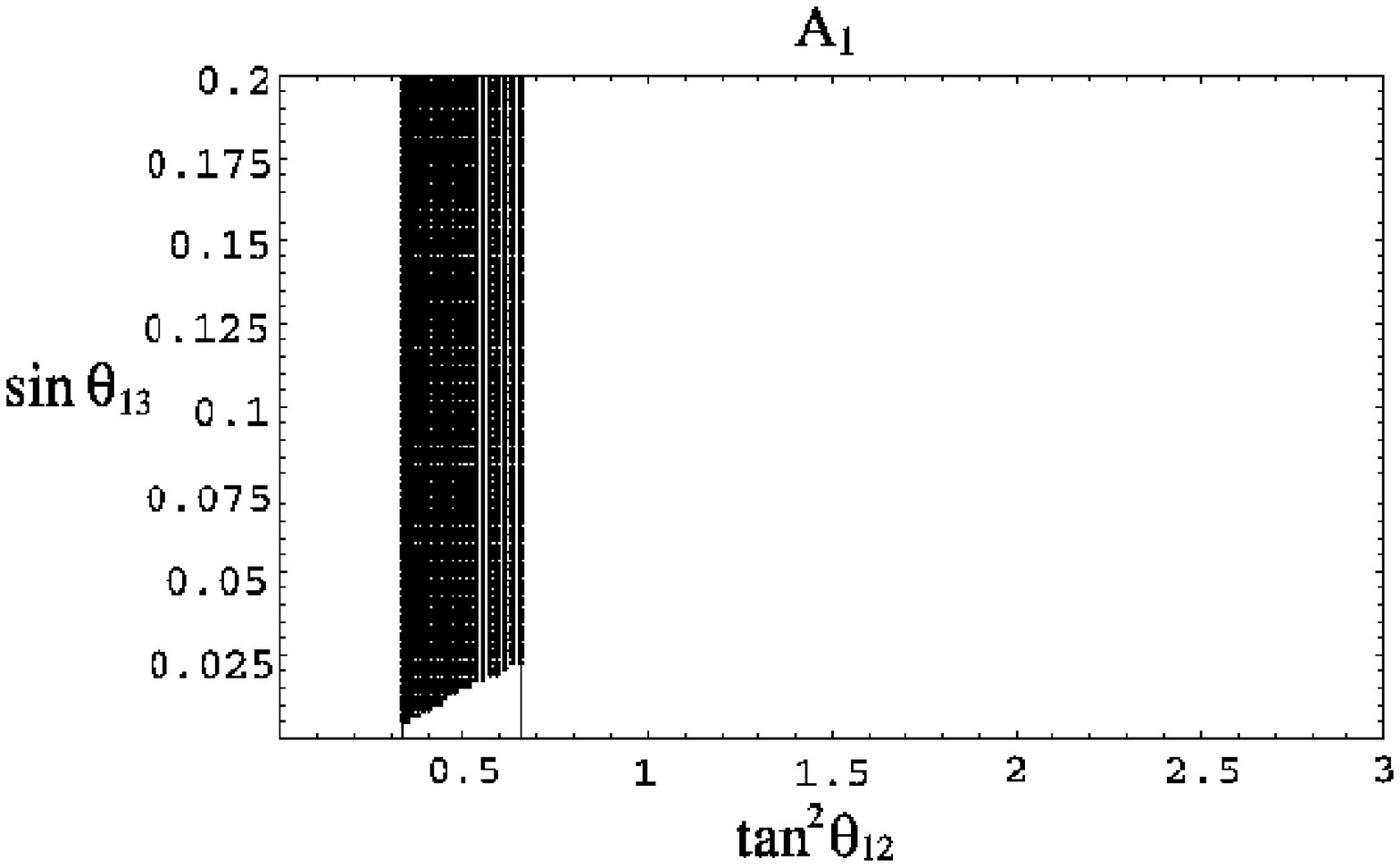}}
\caption{ Scatter plot of $\sin\theta_{13}$ versus $\tan^2\theta_{12}$
  in the case of $\kappa=2\bar\omega=0.07$.
 }
\end{figure}
\begin{figure}
\epsfxsize=7 cm
\centerline{\epsfbox{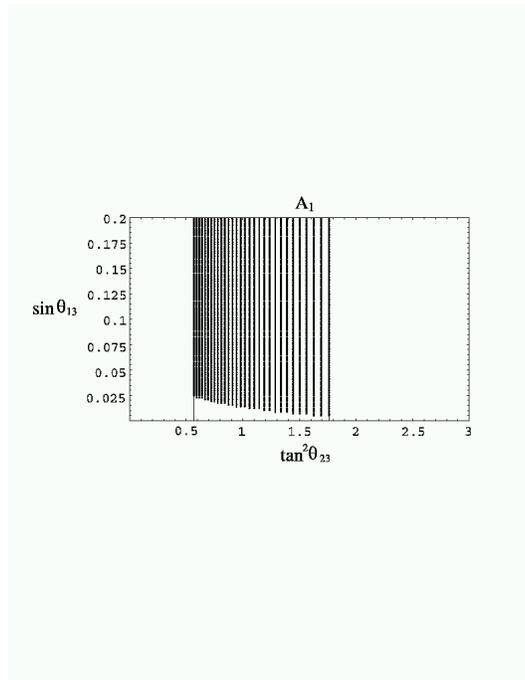}}
\caption{ Scatter plot of $\sin\theta_{13}$ versus $\tan^2\theta_{23}$
 in the case of $\kappa=2\bar\omega=0.07$. }
\end{figure}

In order to estimate the effect of $\bar\omega$ and $\bar\epsilon$,
  we consider the case in which the charged lepton mass matrix has 
 small off-diagonal components. 
Suppose that the  two zeros 
 in eq.(\ref{A12}) is still preserved for the neutrino sector.
The typical model of the charged lepton 
 is the Georgi-Jarlskog texture \cite{GJ},
in which the charged lepton mass matrix $M_E$ is given as
\begin{equation}
M_E \simeq  \ \left ( \matrix{
 0 &\sqrt{m_e m_\mu} & 0 \cr \sqrt{m_e m_\mu} & m_\mu & \sqrt{m_e m_\tau} \cr
 0  & \sqrt{m_e m_\tau} & m_\tau \cr} \right ) \ ,
\label{GJtex}
\end{equation}
\noindent where each matrix element is written in terms of the
charged lepton masses,  and phases are neglected for simplicity.
This matrix is diagonalized by the unitary matrix $U_E$, 
in which the mixing between the first and second families 
is $\sqrt{\frac{m_e}{m_\mu}}\simeq 0.07$ and 
the mixing between the second  and third families 
is $\sqrt{\frac{m_e}{m_\tau}}\simeq 0.02$.
Since the neutrino mass matrix is still the texture $A_1$,
  it turns to
\begin{eqnarray}
M_\nu \sim \ \ \left ( \matrix{ \kappa^2 &\kappa &\lambda \cr
\kappa & 1 & 1 \cr \lambda  & 1 & 1 \cr} \right ) \ ,
\label{A1modi}
\end{eqnarray}
\noindent in the diagonal basis of the charged lepton mass matrix.
Here only the leading mixing term
of $\kappa= \sqrt{\frac{m_e}{m_\mu}}$ is taken.

By using the texture of the neutrinos in  eq.(\ref{A1modi}),
we show our results of  the allowed region of $\sin\theta_{13}$ versus
$\tan^2\theta_{12}$ and $\tan^2\theta_{23}$ in Fig.4 and Fig.5, respectively.
These results should be compared with the ones in Fig.2 and Fig.3.
 It is noticed that the lower bound of $\sin\theta_{13}$ 
considerably comes down due to the correction $\kappa$.
  The small $U_{e3}$ of  $5\times 10^{-3}$ is allowed.
\section{Summary}
We have discussed   $|U_{e3}|$ in the models, in which
the samll  $|U_{e3}|$ is predicted.  The first one is the
model  in which the bi-maximal mixing is realized at the symmetric limit.
The second one is the texture zeros of the neutrino mass matrix.
In the first model,  $|U_{e3}|= 0.03 \sim 0.2$ is predicted.
In the second model,
the lower bound of $|U_{e3}|$
 is given as $0.05$, which considerably depends on  $\tan^2\theta_{12}$ and 
$\tan^2\theta_{23}$. We have investigated the stability of these predictions 
by taking account of small corrections, which may come from radiative 
corrections or off-diagonal elements of the charged lepton mass matrix.  
The lower bound of $|U_{e3}|$ comes down significantly
in the case of  $\bar\omega\gg 0.01$, while $\bar\epsilon$  
is rather insensitive to  $|U_{e3}|$. 
 The measurement of  $|U_{e3}|$ will be an important  test of the texture
zeros  in the future.

\vskip 0.5 cm

 This talk is based on the reserach work with M. Honda and S. Kaneko.
 The research was supported  by  the Grant-in-Aid for Science Research, Ministry of Education, Science and Culture, Japan(No.12047220). 

\end{document}